\documentclass[prl,aps,showpacs,onecolumn,superscriptaddress,amsmath,amssymb,preprint]{revtex4-1}
\usepackage{graphicx}
\usepackage{color}

\usepackage{ulem}
\usepackage{enumitem}

\begin{document}

\title{Physical origin of higher-order soliton fission in nanophotonic semiconductor waveguides}
\author{Charles Ciret}
\affiliation{OPERA-Photonique, Universit\'e libre de Bruxelles (ULB), 50 av. F.D. Roosevelt, CP194/5, B-1050 Bruxelles, Belgium}
\affiliation{ Laboratoire de Photonique d’Angers EA 4464, Université d’Angers, 2 Boulevard Lavoisier, 49000 Angers, France}
\author{Simon-Pierre Gorza}
\affiliation{OPERA-Photonique, Universit\'e libre de Bruxelles (ULB), 50 av. F.D. Roosevelt, CP194/5, B-1050 Bruxelles, Belgium}
\author{Chad Husko}
\affiliation{Center for Nanoscale Materials, Argonne National Laboratory, Argonne, IL 60439}
\author{Gunther Roelkens}
\author{Bart Kuyken}
\affiliation{Photonics Research Group, Department of Information Technology, Ghent University-IMEC, B-9000 Ghent, Belgium}
\author{Fran\c cois Leo}
\email{francleo@ulb.ac.be}
\affiliation{OPERA-Photonique, Universit\'e libre de Bruxelles (ULB), 50 av. F.D. Roosevelt, CP194/5, B-1050 Bruxelles, Belgium}
%





\begin{abstract}

Supercontinuum generation in Kerr media has become a staple of nonlinear optics. It has been celebrated for advancing the understanding of soliton propagation as well as its many applications in a broad range of fields. Coherent spectral broadening of laser light is now commonly performed in laboratories and used in commercial "white light" sources. 
The prospect of miniaturizing the technology is currently driving experiments in different integrated platforms such as semiconductor on insulator waveguides.
Central to the spectral broadening is the concept of higher-order soliton fission. While widely accepted in silica fibers, the dynamics of soliton decay in semiconductor waveguides is yet poorly understood.
In particular, the role of nonlinear loss and free carriers, absent in silica, remains an open question.
Here, through experiments and simulations, we show that nonlinear loss is the dominant perturbation in wire waveguides, while free-carrier dispersion is dominant in photonic crystal waveguides.
\end{abstract}


\maketitle

\section*{Introduction}

Solitons are ubiquitous in science. They correspond to localized packets that propagate unperturbed as a consequence of a balance between nonlinear self-focusing and a diffusion-like process. They have been theoretically and experimentally investigated in hydrodynamics~\cite{john_scott_russell_report_1845}, plasma physics~\cite{kuznetsov_soliton_1986}, biology~\cite{davydov_collective_1985} and optics~\cite{kivshar_chapter_2003}. In the latter, the sech-shaped solutions of the nonlinear Schr\"odinger equation (NLSE), known as solitons, are the most notable~\cite{zakharov_exact_1972}.
Initially considered as potential carriers of information, the focus recently shifted to the central role they play in the process of nonlinear spectral broadening~\cite{skryabin_colloquium_2010}.
NLSE solitons can be of different orders, with only the first order (fundamental) soliton strictly maintaining a constant profile during propagation. Higher-order solutions instead display a periodically varying shape~\cite{mollenauer_experimental_1980}. When perturbed, higher-order solitons tend to split into several fundamental solitons, each centered at a different wavelength, potentially spanning several octaves. This process, commonly called higher-order soliton fission, is the basic mechanism underlying supercontiuum generation technology that finds application in a broad range of fields~\cite{dudley_supercontinuum_2006}.
Initially performed in bulk crystals~\cite{alfano_emission_1970}, supercontinuum generation has been promoted by the advent of photonic crystal fibers~\cite{ranka_visible_2000}. The long interaction lengths and small mode areas facilitated the observation of soliton fission and triggered a wealth of experimental investigations~\cite{dudley_supercontinuum_2006}.
While any perturbation to the NLSE will eventually lead to the decay of a higher-order soliton, it is now well understood that the Raman effect and higher-order dispersion (HOD) govern the dynamics of the fission process in optical fibers.

As there currently is a big interest in the miniaturization of supercontinuum sources, many experimental investigations of spectral broadening through fission of higher-order solitons in nanophotonic waveguides have been reported~\cite{hsieh_supercontinuum_2007,lamont_supercontinuum_2008,ding_time_2010,halir_ultrabroadband_2012,husko_soliton_2013,leo_generation_2014,dave_dispersive-wave-based_2015}. With most of the focus on the spectral broadening, discussions on the dynamics of the fission remain scarce. 
The broadest spectra have been obtained on platforms made of wide band-gap media such as silicon nitride~\cite{guo_mid-infrared_2018} and chalcogenide glasses~\cite{yu_broadband_2014} where the nonlinear loss is low. The dynamics of soliton fission in these waveguides is inferred to be very similar to that of optical fibers. In contrast, soliton fission in narrow band-gap semiconductors such as silicon and indium gallium phosphide (InGaP) is much less understood. The band structure of the medium complexifies the dynamics as nonlinear absorption and free carriers may impact soliton propagation~\cite{yin_dispersion_2006,colman_temporal_2010}. 
Supercontinuum generation through fission of higher-order solitons in semiconductor waveguides was first proposed in 2007~\cite{yin_soliton_2007}. Other theoretical studies followed~\cite{osgood_engineering_2009,husko_soliton_2013,blanco-redondo_observation_2014}. Few however discuss the impact of individual perturbative terms on the soliton dynamics. 

In our recent report on supercontinuum generation in silicon wires~\cite{leo_dispersive_2014} pumped around the 1550~nm telecommunication wavelength, we argued that the main perturbation to the NLSE describing our measurements is two-photon absorption (2PA). Soliton fission provoked by nonlinear loss was shown numerically by Silberberg~\cite{silberberg_solitons_1990}, and in early experimental investigations of one-dimensional spatial soliton propagation in solid-state waveguides~\cite{aitchison_observation_1990,aitchison_spatial_1991}. Temporal and spatial solitons being mathematically equivalent, 2PA is likely to play an important role in the temporal domain as well. Simulations describing our experiments indeed showed a fairly symmetrical post-fission dynamics indicative of 2PA, barring the emission of a weak dispersive wave (DW). 
It is well known that solitons can shed energy when perturbed by HOD~\cite{akhmediev_cherenkov_1995}. 
We inferred when analyzing our data that the weak symmetry breaking we observed was solely due to DW emission. This lead us to the conclusion that 2PA is the main cause for soliton decay in a silicon wire pumped around 1550~nm~\cite{leo_dispersive_2014,leo_coherent_2015}.

In contrast, another recent report claims to experimentally observe free-carrier induced fission of higher-order solitons in InGaP photonic crystal waveguides (PhCWG)~\cite{husko_free-carrier-induced_2016}. In the supplementary section, through analytical investigation of the perturbed NLSE, the authors suggested that free-carrier dispersion (FCD) can also trigger soliton decay in 1550~nm pumped silicon wires. However, the authors did not consider the role of the wire waveguide geometry, nor 2PA interplay with the FCD effect, and the analysis appears incomplete. Free carriers are well known to impact pulse propagation in integrated waveguides~\cite{lin_nonlinear_2007}. The nonlinear loss inherent to semiconductor devices results in the excitation of electron-hole pairs that subsequently impact the dynamics in two different ways. They absorb photons (free carrier absorption, FCA) as well as change their wavenumber (free carrier dispersion, FCD)~\cite{soref_electrooptical_1987}. The latter is the mechanism underlying carrier depletion semiconductor phase and amplitude modulators~\cite{xu_micrometre-scale_2005}. It can also alter short pulse propagation through phase modulation on the leading edge of the pulse and was demonstrated to impact soliton propagation~\cite{monat_slow_2009}. FCA on the other hand has less impact on soliton propagation as the induced loss is low compared to the instantaneous multi-photon absorption processes (see below).

The conclusions of~[\citenum{leo_dispersive_2014}] and~[\citenum{husko_free-carrier-induced_2016}] are difficult to reconcile, leaving the reader wondering what drives the dynamics of ultrashort pulse propagation in semiconductor nanowaveguides. Here, we aim to clarify the role that both nonlinear loss and FCD play on higher-order soliton fission.  We start by investigating the case of pulse propagation in a silicon wire waveguide pumped around the telecommunication wavelength where nonlinear loss is dominated by 2PA. We perform new experiments of supercontinuum generation and confirm that FCD has very little impact on the dynamics on the wire waveguide geometry. Furthermore we theoretically investigate the case of InGaP waveguides. When pumped at 1550~nm, the lowest order nonlinear loss is three-photon absorption (3PA). We review the impact of 3PA on the propagation of a higher-order soliton. We find that nonlinear loss can induce soliton fission in InGaP wire waveguides and that the impact of carriers strongly depends on the waveguide dispersion and input pulse characteristics. From these broad investigations, we derive a general set of conclusions about the roles of nonlinear loss and FCD in higher-order soliton fission in nanophotonic semiconductor waveguides.

\section*{Results}

\textbf{Soliton fission in 2PA-limited silicon nanowires}.
\textit{Experiments and numerical simulations}.
In previous experimental investigations of soliton fission, the main objective was to demonstrate that broad supercontinuum generation can be achieved \textit{despite} strong nonlinear loss~\cite{leo_dispersive_2014,leo_coherent_2015}. They were therefore focused on waveguides whose dispersion properties maximized the spectral broadening. Specifically, it was shown that carefully choosing the waveguide dimensions allows to precisely control the emission of DWs.  In these measurements, DWs were emitted on the blue side of the soliton only, inducing a spectral asymmetry that may be misinterpreted as FCD induced blue shift. 
Here, we report new experiments performed in silicon wire waveguides designed to have notably different dispersion profiles. The objective of these experiment is to further probe the role of carriers on supercontinuum generation in the presence of strong 2PA.

\begin{figure}[ht]
\centering
 \includegraphics[width=15.5cm]{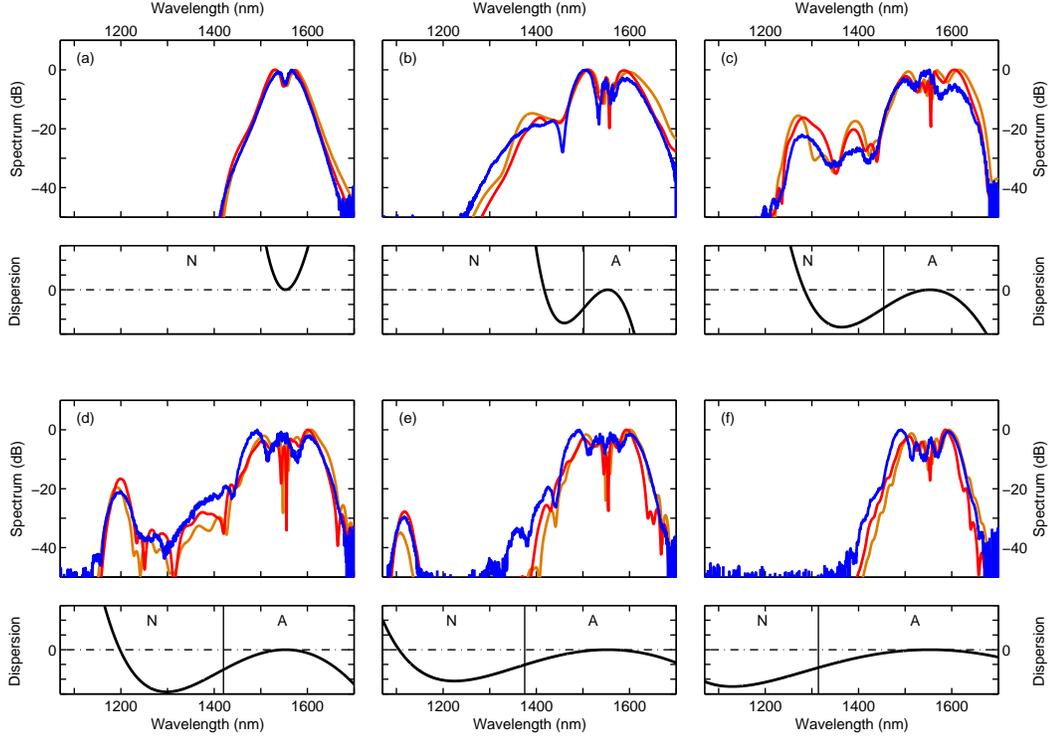}
\caption{\textbf{Experimental demonstration of soliton fission in silicon nanowaveguides.} Top panel of (a)-(f): experimental (blue) spectra measured at the output of the waveguides. The nominal widths are: (a)~800, (b)~700, (c)~675, (d)~650, (e)~625 and (f)~ 575 nm. Also shown are the theoretical spectra computed with equation~\eqref{eq : GNLSE2} (full NLSE, red line) and equation~\eqref{eq:NLSE} (reduced NLSE, brown line). The widths used to compute the dispersion properties differ from the nominal ones, they are: (a)~820, (b)~690, (c)~650, (d)~620, (e)~583 and (f)~ 530 nm.
Bottom panel of (a)-(f): Wavenumber profiles of the waveguide as used in the corresponding simulations shown in the top panel. The vertical continuous line highlights the zero-dispersion wavelength. The normal and anomalous dispersion regions are noted with their respective initial. Different vertical axis were used in each figure for clarity. The ticks spacing respectively correspond to (a)~0.1, (b)~0.1, (c)~1, (d)~2.7, (e)~10 and (f)~26.7 $\times10^{-3}$m$^{-1}.$}
\label{fig:exp}
\end{figure}

We use 6~mm long, 220~nm high silicon-on-insulator wire waveguides. We consider six different waveguides, with widths ranging from 800~nm to 575~nm. We inject 165~fs pulses at 1556~nm with a peak power of 30~W at the input of the nanowires (See methods).
The measured output spectra are shown in Figure~\ref{fig:exp} (blue curves). We readily observe (i) evidence of soliton fission and dispersive wave emission, and (ii) that the broadening highly depends on the waveguide width. In order to gain insight into our experimental results we perform numerical simulations of the generalized nonlinear Schr\"odinger equation (GNLSE) describing ultra-short pulse propagation in silicon nanowires [See Equation~\eqref{eq : GNLSE2} in methods]. The dispersion profile is computed by use of a mode solver (see methods). We stress that calculated dispersion profiles are known to deviate from measured ones. For example, the dispersion of a 220~nm high silicon wire waveguide with a width of 500~nm (as measured in~[\citenum{leo_generation_2014}]) is best reproduced by mode solver simulations when considering a 420~nm wide waveguide with the same height. We account for these deviations by treating the waveguide width as a free parameter in the simulations. Note that all other parameters are fixed.
The computed spectra are shown in Figure~\ref{fig:exp} and are in excellent agreement with the experiments.
Also shown in Figure~\ref{fig:exp} are the dispersion profiles used in simulations. Detailed spatial evolution of the simulated (temporal and spectral) pulse profiles can be found in the supplementary materials. 

The spectrum at the output of the first waveguide [Figure~\ref{fig:exp}(a)] is consistent with a propagation in the normal dispersion regime as predicted by the computed dispersion. The spectral broadening is induced by self-phase modulation only and the pulse temporally spreads as it propagates. 
As the input spectrum moves into the anomalous dispersion regime [Figure~\ref{fig:exp}(b)], we notice an increased but asymmetric spectral broadening. It stems from higher-order soliton compression and subsequent emission of a blue detuned dispersive wave~\cite{leo_dispersive_2014}. The position of the latter is well predicted by the zero crossing of the linear wavenumber as expected from theory~\cite{akhmediev_cherenkov_1995,erkintalo_cascaded_2012}.
As we further reduce the waveguide width, the zero-dispersion wavelength (vertical line) moves away from the pump and the dispersive wave progressively shifts to shorter wavelengths.
In the narrowest waveguide [Figure~\ref{fig:exp}(f)], the phase matching point is located well below the band-gap of silicon ($\approx 1120~$nm) such that the process is hampered.

The excellent agreement between experimental and numerical results constitute a good starting point for discussing the impact of the different perturbations on soliton propagation.
We start by introducing a model solely comprising HOD, linear loss and two-photon absorption as perturbations to the NLSE.
This equation, describing the evolution of the temporal envelope $E(z,t)$ of the electric field along $z$, reads:
\begin{equation}\label{eq:NLSE}
\frac{\partial E}{\partial z} + i\frac{\beta_2}{2}\frac{\partial^2 E}{\partial t^2}-i\gamma |E|^2E = i \sum_{k\ge 3} i^k \frac{\beta_k}{k!}\frac{\partial^k E}{\partial t^k} -\frac{\alpha_\mathrm{l}}{2}E -\alpha_{\mathrm{2PA}} |E|^2E,
\end{equation}
where $\beta_k$ are the Taylor series expansion coefficients of the dispersion profiles shown in Figure~\ref{fig:exp}, $\alpha_\mathrm{l}$ and $\alpha_{\mathrm{2PA}}$ account respectively for the linear and nonlinear losses and $\gamma$ is the nonlinear parameter of the waveguide.
We compare (in Figure~\ref{fig:exp}) the spectra computed with equation~\eqref{eq:NLSE} with those obtained with equation~\eqref{eq : GNLSE2} that includes among others the FCD term. We find only minor differences between the two, hinting that only 2PA and HOD impact pulse propagation (linear loss is negligible over such a short distance). We contend that the agreement validates the use of equation~\eqref{eq:NLSE} for ultra-short pulse propagation in silicon waveguides in the presence of strong 2PA, as predicted in~[\citenum{yin_soliton_2007}].
Whether it is 2PA or HOD that dominates is likely to be sensitive to the experimental conditions. In optical fibers for example, Raman and HOD introduce comparable perturbations and which induces the fission highly depends on input pulse duration~\cite{dudley_supercontinuum_2006}. While we found that 2PA and HOD suffice to explain our results, it seems pertinent to investigate the impact of each perturbation independently in an effort to gain further physical insight into the dynamics of soliton fission in semiconductor waveguides. Before considering each perturbation independently, we will consider the full GNLSE equation~\eqref{eq : GNLSE2} for the sake of completeness.

\begin{figure}[h]
\centering
\includegraphics[width=15.5cm]{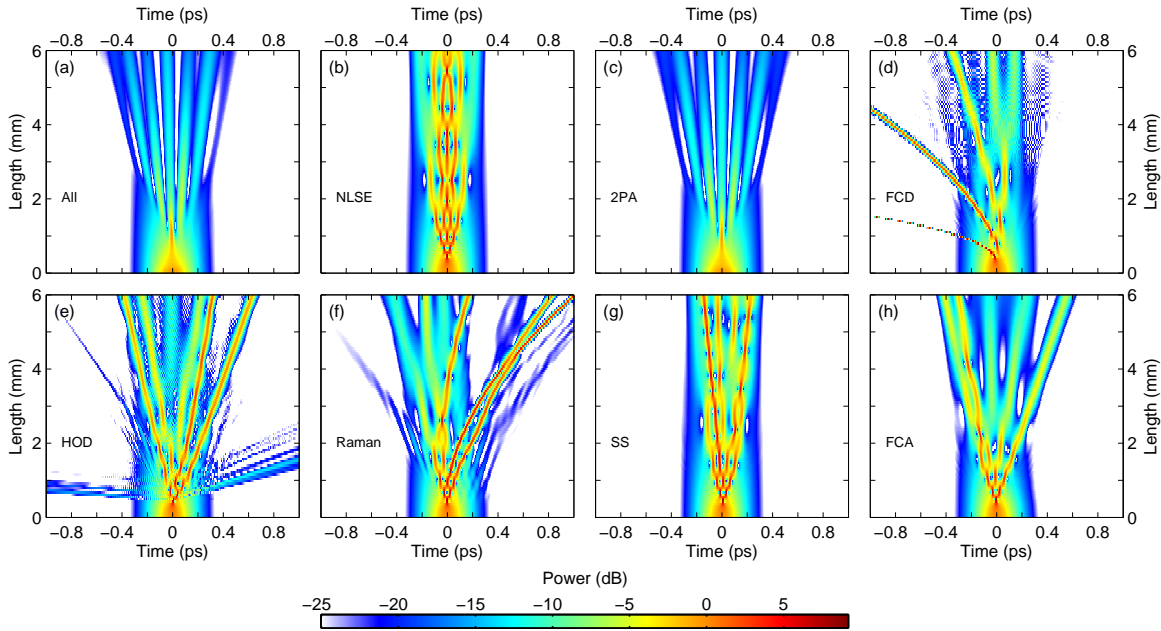}
\caption{\textbf{Spatial evolution of the temporal pulse profile in a silicon waveguide.} Comparison of the propagation when computed with (a)~the GNLSE [equation~\eqref{eq : GNLSE2}],(b)~the NLSE, (c)~the NLSE and 2PA,(d)~the NLSE and FCD, (e)~the NLSE and HOD, (f)~the NLSE and the Raman effect,(g)~the NLSE and self-steepening, (h)~the NLSE and FCA. Note that linear loss is included in all cases. The colormap is normalized with respect to the input peak power.}
\label{fig:fission}
\end{figure}

For the full GNLSE simulations, we focus on the dispersion profile corresponding to the narrowest waveguide where no DW is emitted to clearly discriminate between the impact of HOD on the fission and DW emission [see Figure~\ref{fig:exp}(f)].
The temporal profile evolution as predicted by equation~\eqref{eq : GNLSE2} is shown in Figure~\ref{fig:fission}(a). The propagation is reminiscent of the higher-order soliton fission dynamics observed in optical fibers, with signatures of (i) an early temporal compression stage and (ii) subsequent splitting into several pulses. 
We note that due to the strong nonlinear loss, it is not clear if these pulses qualify as solitons. It was suggested that they can be described using the concept of path-averaged solitons~\cite{yin_dispersion_2006}. 
In contrast, when nonlinear loss is absent (as discussed below), we have confirmed that the ejected pulses are fundamental solitons.
A detailed analysis of the propagation of fundamental solitons in the presence of 2PA is outside the scope of the present work.
Nevertheless, a striking difference, as compared to the dynamics of higher-order solitons in optical fibers, is that the propagation here is quite symmetrical in the time domain, while most of the perturbations (FCD, HOD, Raman, self-steepening) break the time-reversal symmetry of the NLSE. For example, self-frequency shift as induced by the Raman effect in silica fibers is notably absent. We discern however a slight acceleration of the pulses, as seen from the left leaning pulse trajectory in Figure~\ref{fig:fission}(a), that is induced by the carriers~\cite{blanco-redondo_observation_2014}.

We next consider the evolution of the higher-order soliton profile under the separate action of each perturbation~[Figure~\ref{fig:fission}~(c)-(h)], as well as without perturbation ~[Figure~\ref{fig:fission}~(b)]. Linear loss is systematically included in the simulations but we note that while it may induce fission on its own (see e.g.~[\citenum{prilepsky_breakup_2007}]), we have not observed such behavior over the length of the sample.
In Figure~\ref{fig:fission}(b), we display the dynamics of the NLSE solely perturbed by linear loss. It is reminiscent of the propagation of a higher-order soliton. We checked that the propagation is periodic in the absence of loss. The period coincidentally corresponds to the length of the waveguide.
When we introduce 2PA in the simulations [Figure~\ref{fig:fission}(c)], we observe a purely symmetrical fission that looks strikingly similar to that computed with the full GNLSE [Figure~\ref{fig:fission}(a)], suggesting that 2PA is the dominant effect in the current silicon wire waveguide geometry.
%
\\
While it is apparent that we can describe this current device with only 2PA, we continue our examination of the different perturbations. If we include FCD only [Figure~\ref{fig:fission}(d)], the dynamics is very different. It is then governed by soliton acceleration as a consequence of carrier-induced blue shift. This is consistent with the dynamics simulated in~[\citenum{husko_free-carrier-induced_2016}] and confirms that the small acceleration observed in Figure~\ref{fig:fission}(a) is due to the presence of carriers. 
In Figure~\ref{fig:fission}(e) we show the impact of HOD. It is well known that HOD may both provoke soliton decay and lead to DW emission~\cite{dudley_supercontinuum_2006}. Remember that we purposely investigate the case where no DW is emitted in the experiment to be able to discriminate between the two. 
Without nonlinear loss however, the spectrum of the higher-order soliton is much broader at the point of compression such that it emits DW's in the normal dispersion regime on both sides of the pump. We expect that the dynamics shown in Figure~\ref{fig:fission}(e) captures the main features of higher-order soliton propagation in waveguides with negligible nonlinear loss and Raman effect such as SiN-on-insulator nanowires~\cite{halir_ultrabroadband_2012,zhao_visible--near-infrared_2015}. In silicon however, we see no sign of the impact of HOD on the propagation outside of DW emission.
\\
\indent
The fission with Raman only is shown in Figure~\ref{fig:fission}(f). The propagation looks similar to the dynamics reported in optical fibers~\cite{dudley_supercontinuum_2006} where the Raman effect induces soliton deceleration by continuously red-shifting solitons as they propagate.
We note however that the Raman gain of silicon consists of a 105~GHz wide Lorentzian centered around 15~THz~[\citenum{temple_multiphonon_1973,lin_nonlinear_2007}] and is thus very different from the one of silica glass. Most ejected solitons are in this case spectrally broad enough for intra-pulse stimulated Raman scattering to take place. In contrast, the ones emitted in simulations of the GNLSE are spectrally narrower (because of the nonlinear loss) such that Raman self-frequency shift does not occur.
Self-steepening (i.e. the shock term) is also capable of inducing fission [Figure~\ref{fig:fission}(g)] as was previously studied in the case of optical fibers~\cite{golovchenko_decay_1985}. Here it seems to have a very limited effect. It is the only perturbation that does not induce soliton ejection at the first compression point when considered independently.
Finally, we see in Figure~\ref{fig:fission}(h) that FCA induced fission looks similar to the dynamics in the presence of 2PA only [Figure~\ref{fig:fission}(b)], however, not enough carriers accumulate over 165~fs to compete with 2PA induced loss. 
In brief, it is apparent that many of these perturbations could induce fission independently. In our case, however, the simulations indicate that the only perturbation to have a significant impact on the dynamics is 2PA. While in the above discussion we focused on a single set of parameters, we note that this conclusion holds for the other cases shown in Fig.~\ref{fig:exp} and in 2PA-limited wire waveguides.

\vspace{0.5cm}
\noindent\textit{Dimensionless analysis}
Given the strong 2PA in our system, we now examine whether FCD can be the main perturbation under different conditions. We here focus on the fission dynamics under the combined action of 2PA and FCD. We start by formulating a normalized model that accounts for both effects. All other perturbations are omitted. Moreover we neglect carrier recombination as the corresponding lifetime is much longer than the input pulse duration.
The equation describing the normalized field envelope $A(\zeta,\tau)$ evolution during propagation (in the anomalous dispersion regime) then reads: 
\begin{equation}\label{eq:normNlSE2}
i\frac{\partial A}{\partial\zeta} + \frac{1}{2}\frac{\partial^2 A}{\partial\tau^2}+ N^2|A|^2A = -N^2\frac{\alpha_{\mathrm{2PA}}}{\gamma}\left[i|A|^2-\eta\int_{-\infty}^{\tau}{|A|^4}d\tau'\right]A,  \end{equation}
where $\zeta$ and $\tau$ are the normalized distance and time parameters (see methods). 
The terms on the left hand side correspond to the standard NLSE. On the right hand side are the 2PA and FCD perturbations. The latter is proportional to the normalized parameter $\eta$. It is equal to the ratio of the 2PA and FCD lengths (as defined in ~\citenum{husko_free-carrier-induced_2016}) and only depends on the free-carrier index change coefficient $k_c$, the effective mode area, and the input pulse energy (see Methods). For a fixed material, the higher the effective fluence [Jm$^{-2}$] in the waveguide, the more impact carriers will have. At this point we note that our normalized equation differs from the one used in~[\citenum{husko_free-carrier-induced_2016}] where 2PA was neglected in order to focus the analysis solely on FCD. 

\begin{figure}[h]
\centering
\includegraphics[width=15.5cm]{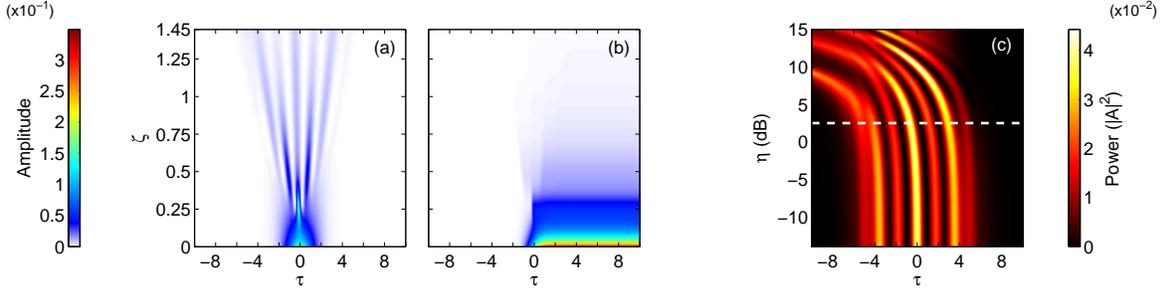}
\caption{\textbf{Impact of 2PA and FCD on pulse propagation}.  (a): Amplitude of the normalized nonlinear loss contribution to the propagation as a function of propagation distance ($\eta=1.78$). (b): Amplitude of the normalized carrier induced phase shift term as a function of propagation distance ($\eta=1.78$). (c) Temporal output profile for different values of the normalized FCD parameter. 
The temporal profiles are computed by use of the normalized model with dominant 2PA~\eqref{eq:normNlSE2}. The dotted line highlights the value corresponding to the experiment reported in Figure~\ref{fig:exp} ($N=6$, $\alpha_{\mathrm{2PA}}=58$ W$^{-1}$m$^{-1}$, $\gamma$ = 323 W$^{-1}$m$^{-1}$, $\beta_2=~-2.12$ps$^2$/m).}
\label{fig:norm}
\end{figure}

2PA and FCD effects are difficult to compare as they impact pulse propagation in very different ways. The former results in instantaneous, power-dependent loss while the latter produces a non-instantaneous index variation. 
We infer however that the normalized amplitude of the corresponding term in equation~\eqref{eq:normNlSE2} is likely to be a good indicator of FCD impact. We hence compare the strength of both perturbations as a function of time ($\tau$) and distance $(\zeta)$. We consider the parameters of the experiment shown in Figure~\ref{fig:exp}(f). The input pulse corresponds to a soliton of order $N=6$. We integrate equation~\eqref{eq:normNlSE2} over the waveguide length [starting from $A(0,\tau)=\,\mathrm{sech}(\tau)$] and plot, in Figure~\ref{fig:norm}, the evolution of the amplitude of the 2PA (a) and FCD (b) perturbations.
We readily note that both are strongly dependent on time and propagation distance. While FCD is stronger than 2PA at the beginning of the propagation, it decays more rapidly as the pulse advances. 
We believe that this is the reason we hardly see any impact of the carriers on the dynamics in our current experiments.
Similarly, using characteristic length scales in the presence of nonlinear loss can be misleading since they are calculated based on input parameters only, and not dynamic interaction of the nonlinearities. 

Next we investigate the potential impact of carriers on soliton propagation. Specifically, we compute, by use of equation~\eqref{eq:normNlSE2}, the temporal pattern at the output of the waveguide for different values of the normalized FCD parameter $\eta$. Our results are shown in Figure~\ref{fig:norm}(c). 
The dotted line indicates the value computed with our experimental parameters ($\eta=1.78$).
The plot starts around $\eta=0.05$ and displays a fully time symmetric output pattern. The pulse propagation dynamics in this case are the same as shown in Figure~\ref{fig:fission}(c), with the symmetric temporal profile unambiguously indicating that the fission dynamics here are driven by 2PA. As we increase $\eta$, we observe a time shift of the pattern that is a signature of carrier induced acceleration. 
Around the dotted line, the very small deviation from the symmetric profile highlights the minimal impact of the carriers in our experiments. Almost the same patterns are found whether we include the carriers or not.

\begin{figure}[t]
\centering
\includegraphics[width=13cm]{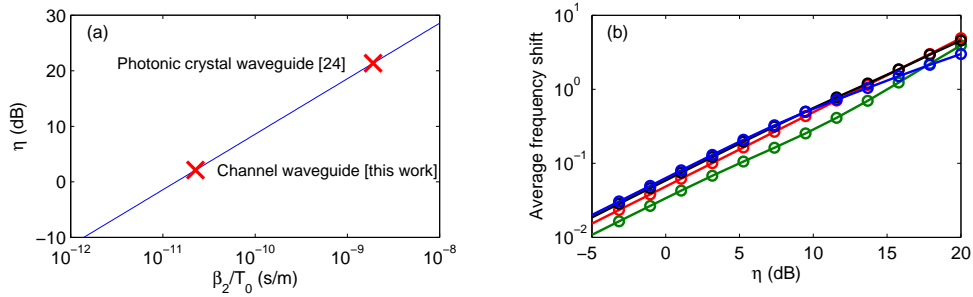}
\caption{\textbf{Analysis of the spectral assymetry}. (a) Normalized parameter $\eta$, corresponding to a silicon waveguide pumped around 1550\,nm with a sixth-order soliton, as a function of the $|\beta_2|/{T_0}$ ratio. The red markers highlight the values corresponding to our experiment in channel waveguides as well as recent experiments in photonic crystal waveguides~\cite{blanco-redondo_observation_2014}. (b) Average frequency shift as a function $\eta$ for different values of the soliton number. $N = 4$ (blue line), $N= 10$ (black line), $N = 20$ (red line) and $N=60$ (green line). Note that a fully symmetric spectrum has a null average frequency.}
\label{fig:eta}
\end{figure}

A numerical increase of the FCD parameter, however, leads to a very different behavior. Not only do the pulses travel faster, as expected from the stronger carrier-induced blue shift, but the pattern also completely loses its time-reversal symmetry. A map of the propagation of the pattern across the waveguide in the case of very strong FCD ($\eta=100$) can be found in the supplementary materials. The post-fission dynamics is reminiscent of the one shown in Figure~\ref{fig:fission}(d) (FCD perturbation only) where solitons of different peak powers are emitted at different times. 
The propagation is in this case mostly governed by carriers such that the fission likely qualifies as "FCD induced". 
We stress however that the physical parameters corresponding to $\eta =100$  do not coincide with those characterizing our experiments in wire waveguides, past or present. They would however describe experiments in silicon photonic crystals~\cite{blanco-redondo_observation_2014}. This can be more easily understood by rewriting the dimensionless FCD parameter as 
\begin{equation}\label{eq:eta}
\eta = \frac{k_c}{\hbar\omega_0n_2}N^2\frac{|\beta_2|}{T_0}
\end{equation}
where the constants common to silicon waveguides pumped around 1550~nm and the soliton number clearly appear, leaving the $|\beta_2|/{T_0}$ ratio as an important parameter allowing to predict the asymmetry.
This is illustrated in Figure~\ref{fig:eta}(a) where we plot $\eta$ as a function of the $|\beta_2|/{T_0}$ ratio for a silicon waveguide pumped around 1550\,nm with a sixth-order soliton.
As discussed above, our experiments correspond to $\eta\approx2$ and display a mostly symmetrical output wave. In photonic crystal waveguides however, the group-velocity dispersion is substantally larger because slow propagation arises close to the photonic band edges. Despite the longer pulses, these experiments are characterized by a $|\beta_2|/{T_0}$ ratio two orders of magnitude larger than for channel waveguides. The lack of demonstration of FCD induced soliton fission in silicon photonic crystals is likely due to the strong linear loss~\cite{blanco-redondo_observation_2014}.
To further stress how important the dimensionless parameter $\eta$ is for predicting the dynamics, we performed simulations of equation~\ref{eq:normNlSE2} for different soliton numbers and different values of $\eta$.
To characterize the asymmetry, we use the average frequency shift as defined by the first moment $\omega_m=\int\omega S(\omega)d\omega$ where $S(\omega)$ is the spectral amplitude of the output wave. Our choice to describe the asymmetry in the Fourier domain is motivated by the fact that spectrum does not change noticeably post-fission (See e.g. Figure S1).
Our results in Figure~\ref{fig:eta}(b) show that the soliton number has very little impact on the spectral asymmetry of the output wave confirming that the normalized parameter $\eta$, proportional to the input energy, is the most relevant parameter to determine the dominant perturbation mechanism.

\vspace{0.5cm}
\noindent\textbf{Soliton fission in 3PA-limited InGaP photonic crystal and wire waveguides}. \textit{Numerical simulations}. We now turn our attention to the dynamics of soliton fission in InGaP waveguides as recently reported in~[\citenum{husko_free-carrier-induced_2016}].
The experiments are performed by pumping a 1.5\,mm long, air-clad, InGaP photonic crystal waveguide. The input pulses are 2.2~ps long (full-width-at-half-maximum), with a peak power of 5.9 W and centered at 1553~nm.
The band gap of InGaP is 1.9 eV such that 3PA is the lowest order nonlinear loss when working around the 1550~nm telecommunication wavelength. Instantaneous nonlinear loss is consequently lower than in silicon.
The experimental results show clear evidence of a second-order soliton decaying into a couple of pulses. 
Through simulations and an analytic expression, the authors argue that it is FCD that induces the fission. Given our experiments above, however, one wonders about the role of 3PA in the dynamics. Here we try to clarify this role by theoretically exploring pulse propagation in an InGaP waveguide.
Specifically we use the same approach as above and simulate the dynamics under the action of each perturbation separately. We consider here only 3PA, FCD and HOD. Linear loss is always included. All the parameters used in our simulations of InGaP photonic crystal waveguides are summarized in table~\ref{table : InGaP}.

\begin{figure}[b]
\centering
\includegraphics[width=15.5cm]{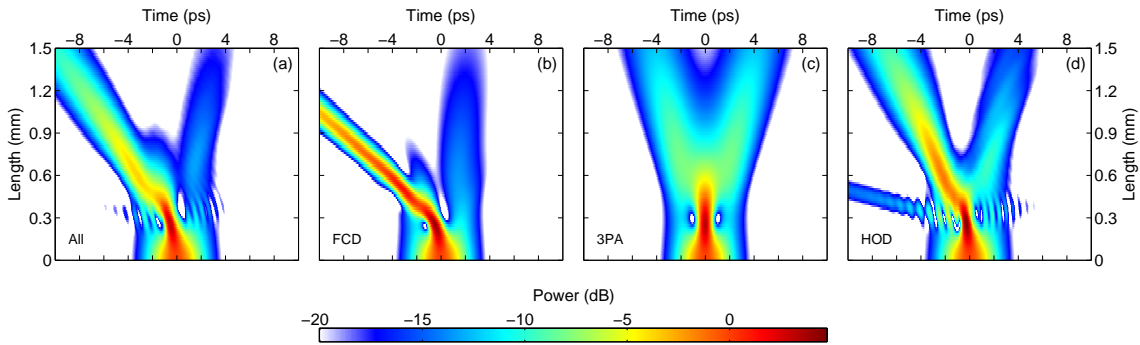}
\caption{\textbf{Spatial evolution of the temporal pulse profile in an InGaP photonic crystal waveguide.} Comparison of the propagation when computed with (a)~the GNLSE [equation~\eqref{eq : GNLSE3}],(b)~the NLSE and FCD, (c)~the NLSE and 3PA, and (d)~the NLSE and HOD. Linear loss is included in all cases. The colormap is normalized with respect to the input peak power.}
\label{fig:fissionInGaP}
\end{figure}

In Figure~\ref{fig:fissionInGaP}(a) we show a map of the propagation through the waveguide when simulated by the GNLSE. Note that it is a different GNLSE than the one used for simulating pulse propagation in a silicon waveguide (see~[\citenum{husko_free-carrier-induced_2016}] and methods for more details).
We observe a simultaneous acceleration and compression of the input soliton followed by a decay into two separate pulses. Dispersive waves are emitted at the point of maximum compression. Next we simulate the dynamics with FCD as the lone perturbation. The temporal profile is shown in Figure~\ref{fig:fissionInGaP}(b). We see both a split of the input soliton and acceleration of the pulses. However, the lack of nonlinear loss noticeably affects the temporal trajectory of the pulses. The increased carrier-induced blue shift causes the first ejected soliton to propagate much faster. 
The dynamics when we take only 3PA into account is displayed in Figure~\ref{fig:fissionInGaP}(c). Interestingly, we remark that the higher-order soliton still decays.  
This shows how 3PA, just as 2PA, impacts the dynamics beyond simply transferring energy from the optical field to the carriers. As expected, the temporal profile fully satisfies time reversal symmetry throughout propagation. 
Finally, in Figure~\ref{fig:fissionInGaP}(d) we show the dynamics when HOD is the lone perturbation to the NLSE. Again, fission occurs.
The case with TOD only was already discussed in~[\citenum{husko_free-carrier-induced_2016}] and displays no fission at all for the low TOD values in that experiment. 
The emission of dispersive waves on both sides of the pump is also an indication that FOD is the main linear perturbation~\cite{mussot_tailoring_2007}.
Importantly, we find that all three perturbations are capable of inducing fission, similar to the 2PA case above. Looking at Figure~\ref{fig:fissionInGaP} it seems likely that all play a part, and we hence further investigate the role of 3PA and in particular how it compares with FCD. 

\begin{figure}[t]
\centering
\includegraphics[width=15.5cm]{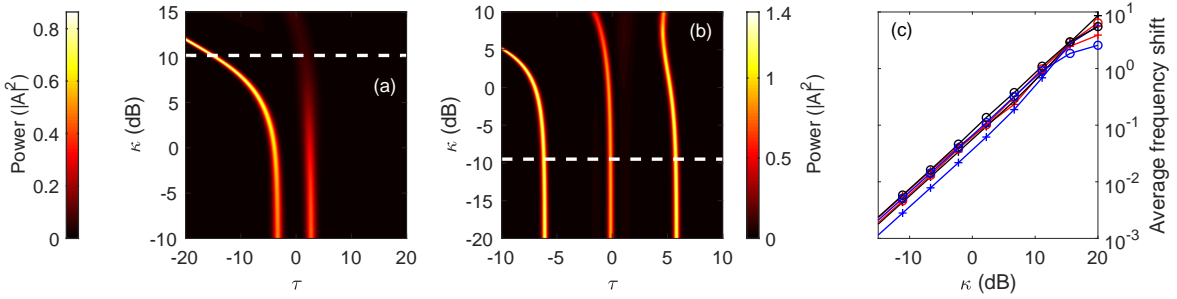}
\caption{\textbf{Impact of FCD on soliton propagation} Temporal output profile for different values of the normalized FCD parameter in the case of (a) an InGaP photonic crystal waveguide ($N=2$ and $\Gamma=2\times10^{-2}$) and (b) an InGaP wire waveguide ($N=4$ and $\Gamma=1.4\times10^{-3}$). The profiles are computed by use of the normalized model with dominant 3PA \eqref{eq:normNlSE3}.
The dotted line highlights the value corresponding to the experiments reported in~[\citenum{husko_free-carrier-induced_2016}] ($\kappa=10.5$) and [\citenum{dave_dispersive-wave-based_2015}] ($\kappa=0.11$). (c) Average frequency shift as a function $\kappa$ for different values of $\Gamma$ and soliton number. $\Gamma = 2\times10^{-3}$ (crosses) and $\Gamma=2\times10^{-2}$ (circles). $N = 4$ (blue line), $N= 10$ (black line) and $N = 20$ (red line). Note that a fully symmetric spectrum has a null average frequency.}
\label{fig:normInGaP}
\end{figure}

\vspace{0.5cm}
\noindent\textit{Dimensionless analysis}
Similarly for 2PA, we introduce a normalized model including only nonlinear loss and FCD: 
\begin{equation}\label{eq:normNlSE3}
i\frac{\partial A}{\partial\zeta} + \frac{1}{2}\frac{\partial^2 A}{\partial\tau^2}+ N^2|A|^2A =
-\Gamma N^2 \left[i|A|^4-\kappa\int_{-\infty}^{\tau}{|A|^6}d\tau'\right]A .
\end{equation}
The perturbation terms on the right hand side are different than in equation~\eqref{eq:normNlSE2} describing propagation in silicon waveguides. They still correspond to nonlinear loss and FCD but scale differently as three photons are now required to generate an electron-hole pair.
We calculate, by integrating equation~\eqref{eq:normNlSE3}, the temporal profile at the output of the waveguide for different values of $\kappa$. We start with the normalized parameters $N=2$ and $\Gamma = 2\times 10^{-2}$ computed from the physical constants used in the simulations shown in Figure~\ref{fig:fissionInGaP}.
The results are shown in Figure~\ref{fig:normInGaP}(a). 
For very low values of the FCD parameter $\kappa$, the output profile is the same as in Figure~\ref{fig:fissionInGaP}(c). Even when the impact of carriers is negligible, the input pulse splits at the first point of compression. The fission is induced by 3PA for these low $\kappa$ values.
As we increase $\kappa$, the carrier-induced blue shift breaks the time reversal symmetry of the pattern.
Most of the energy moves to the front pulse and the distance between the two pulses increases.
The value of $\kappa$ corresponding to the simulations shown in Figure~\ref{fig:fissionInGaP} is highlighted by the dotted line.
Such an important difference indicates that it is indeed FCD that drives the fission process in these experiments in photonic crystal waveguides. We stress that the normalized variables are highly dependent on the waveguide and input pulse characteristics (see methods).

We next probe the impact of carriers in a different geometrical configuration involving InGaP wire waveguides. Specifically, we simulate the propagation in waveguides similar to those recently used for octave-spanning supercontinuum generation~\cite{dave_dispersive-wave-based_2015}. The physical parameters can be found in the methods. The corresponding nonlinear absorption parameter is $\Gamma = 1.4\times 10^{-3}$. We use the same normalized distance as before, corresponding to 4.5 dispersion lengths but we set the input soliton order to $N=4$ because soliton fission does not occur for lower orders over this distance. We vary $\kappa$ keeping $\Gamma$ fixed and integrate equation~\eqref{eq:normNlSE3} over the waveguide length. The results are shown in Figure~\ref{fig:normInGaP}(b).
The output profile is still very symmetrical for the physical FCD value (dotted line). A ten fold increase of $\kappa$ is required for the carriers to noticeably impact the propagation. 
In contrast to the photonic crystal case above, these simulations suggest that 3PA is likely to be the main driver of soliton fission in InGaP wire waveguides. 
Similar to the above analysis for the case of 2PA driven silicon waveguides, we now study the impact of the normalized parameters (here $\kappa$, $\Gamma$ and $N$) on the spectral asymmetry of the output wave.
Our results, shown in Figure~\ref{fig:normInGaP}(c), again emphasize the importance of the FCD parameter as opposed to $\Gamma$ and N that appear to have very little impact on the asymmetry.

\section*{Discussion}

We unveiled the physics driving soliton fission in semiconductor waveguides. This research was motivated by recent results in the literature, each claiming different physical origin: 2PA~\cite{leo_dispersive_2014} or FCD~\cite{husko_free-carrier-induced_2016}. Through a series of experiments and accompanying analysis, we showed that the impact of 2PA and 3PA on ultra-short pulse propagation goes beyond energy loss and can induce fission. We further showed that either 2PA or FCD can be the dominant fission mechanisms depending on the experimental parameters. Our dimensionless analysis suggest that the normalized FCD parameters ($\eta$ for 2PA driven waveguides, $\kappa$ for 3PA driven waveguides) may be used to identify the dominant perturbation to the propagation of higher order solitons in semiconductor waveguides. We found that nonlinear loss (2PA/3PA) are the dominant perturbations in wire waveguides, while FCD is dominant in photonic crystal waveguides. 
We opened this discussion considering the important role of soliton fission in supercontinuum generation. As we often desire supercontinua? with a very broad bandwidth, it is much more likely that wire waveguides will be the primary geometry of interest. A key implication of our work is that nonlinear loss is the main driving mechanism in these systems. The main difference with fission induced by other known perturbations, such as the Raman effect or HOD, is that fission induced by nonlinear loss exhibits symmetric spectral and temporal evolution throughout propagation.

\section*{Methods}

\textbf{Experiment in silicon waveguides}
We use air-clad silicon-on-insulator wire waveguides fabricated in a CMOS pilot line, using 200~mm wafers consisting of a 220~nm silicon layer on top of a 2~$\mu$m buried oxide layer. 
We launch 165~fs (full-width-at-half-maximum) pulses from an OPO (Spectra physics OPAL) pumped by a titanium-sapphire laser (Spectra physics Tsunami) running at 720~nm.  The repetition rate is 82\, MHz. The pulse width was retrieved through autocorrelation measurements. The 1556~nm idler output of the OPO is horizontally polarized and is used to excite the quasi-TE mode of the waveguide.
The light is coupled in the waveguide through a microscope objective (x60, NA = 0.65) and out-coupled by use of a lensed fiber (working distance$\,=14\,\mu$m, NA$\,=0.4$). The respective coupling efficiencies are 23~dB and 7~dB.
The  output signal is measured with an optical spectrum analyzer (Yokogawa AQ 6370D).

\noindent\textbf{Modeling of pulse propagation in silicon waveguides}
We use the following generalized nonlinear Schr\"odinger equation to simulate pulse propagation in the waveguides~\cite{yin_soliton_2007}.
\begin{multline}\label{eq : GNLSE2}
\frac{\partial E(z,t)}{\partial z} = \mathcal{F}^{-1}\left[iD(\omega)\tilde{E}(z,\omega)\right]
 -\frac{\alpha_\mathrm{l}}{2}E -\alpha_{\mathrm{2PA}} |E|^2E- \left(\frac{\sigma}{2} + ik_0k_c\right)N_cE \\ + i\gamma\left(1+\frac{i}{\omega_0}\frac{\partial }{\partial t}\right)E\int^t_{-\infty} R(t-t')|E(z,t')|^2 dt',
\end{multline}
where $E(z,t)$ describes the slowly varying envelope of the field as a function of propagation distance $z$ and time $t$ and
$\tilde{E}(z,\omega)=\mathcal{F}\left[E(z,t)\right]$ is the Fourier transform of the field.  $D(\omega) = \left(\beta (\omega)-\beta (\omega_0)-\left.\frac{\partial\beta}{\partial\omega}\right|_{\omega_0}(\omega-\omega_0)\right)$ is the dispersion operator where $\beta(\omega)$ is the frequency dependent wave vector and $\omega_0$ denotes the pump frequency. $\beta(\omega)$ is computed by use of a mode solver (Lumerical). The data is available on request. $\alpha_\mathrm{l}$ characterizes linear loss. It was evaluated at 2~dB/cm by cutback measurements on similar waveguides. $\sigma N_c$ correspond to the free carrier induced loss where $\sigma = 1.45\times
10^{-21}\,\mathrm{m}^2$~[\citenum{lin_nonlinear_2007}] and 
$N_c$ is the carrier density, computed by solving:
\begin{equation}\label{Eq : carriers2}
  \frac{\partial N_\mathrm{c}(z,t)}{\partial t} =
    \frac{\alpha_{\mathrm{2PA}}}{\hbar\omega_0 A_\mathrm{3eff}}|E(z,t)|^4-\frac{N_\mathrm{c}(z,t)}{\tau_\mathrm{c}}\,,
\end{equation}
where $A_\mathrm{3eff}$ is the effective area related to a third order process, computed with the mode solver.
The nonlinear parameter $\gamma = n_2\omega_0/(cA_\mathrm{3eff})$ and 2PA coefficient $\alpha_{\mathrm{2PA}} = \beta_{\mathrm{2PA}}/(2A_\mathrm{3eff})$ are extracted from measurements on similar waveguides~\cite{matres_high_2013} and scaled through the effective mode area. As the latter is different for each waveguide, so are the nonlinear parameter and 2PA coefficient. The values we used are listed in table~\ref{table : Si}.
$k_0k_cN_c$ is the wavenumber change induced by FCD where $k_0=\omega_0/c$ and $k_c= (8.8\times10^{-28}\,\mathrm{m}^3\bar{N_c} + 1.35\times 10^{-22}\,\mathrm{m}^3\bar{N_c}^{0.8})/\bar{N_c}$ with $\bar{N_c} = N_c\times 1$m$^{3}$.
We stress that $k_c$ does not remain constant during propagation and must be evaluated at each step along with the carrier density. $R(t) = (1-f_\mathrm{R})\delta(t) +
f_\mathrm{R}h_\mathrm{R}(t)$ is the delayed nonlinear function where $f_\mathrm{R}$ is the fractional Raman contribution and $h_\mathrm{R}(t)$ is the
Raman response function. They are deduced from the spectral Lorentzian shape of the Raman response of
silicon~\cite{lin_nonlinear_2007}. We use the relation $f_\mathrm{R} = g_\mathrm{R}(\omega_0)
\Gamma_\mathrm{R}/[\Omega_\mathrm{R}A_\mathrm{eff}\gamma]$ with $g_\mathrm{R}(\omega_0) =3.7\times
10^{-10}\,\mathrm{m/W}$~[\citenum{Claps_Observation_2003}], $\Omega_\mathrm{R}/(2\pi)=15.6$\,THz and
$\Gamma_\mathrm{R}/\pi=105$\,GHz~[\citenum{lin_nonlinear_2007}].
The normalized equation~\eqref{eq:normNlSE2} is linked to equation~\eqref{eq : GNLSE2} through the following relations:

\begin{table}[h]
\centering
\begin{tabular}{l l l l l l}
$N^2 = \frac{L_\mathrm{D}}{L_\mathrm{NL}}$ & $A = \frac{E}{\sqrt{P_0}}$ & $\zeta = \frac{z}{L_\mathrm{D}}$ &  $\tau = \frac{t}{T_0}$ & $\eta = \frac{k_ck_0P_0T_0}{\hbar\omega_0A_{\mathrm{3eff}}}$ \\
\end{tabular}
\end{table}
\noindent where  $L_\mathrm{D} = {T_0^2}/{|\beta_2|}$ and $L_\mathrm{NL}= {1}/{(\gamma_RP_0)}$ are the dispersion and nonlinear lengths as defined at the input of the waveguide, considering an input pulse $\sqrt[]{P_0}\mathrm{sech}(t/T_0)$. 
Note that for the normalized model, HOD, FCA, self-steepening as well as the Raman effect are neglected. Moreover we consider that the FCD parameter $k_c$ remains constant throughout the waveguide. We use $k_c = 3.8\times10^{-27}$ m$^{3}$, computed with the input pulse parameters.

\begin{table}[h]
\centering
\begin{tabular}{l l l l l}
Width [nm] & $A_{\mathrm{3eff}}$ [$\mu$m$^2$]& $\gamma$ [(Wm)$^{-1}$] & $\alpha_{\mathrm{2PA}}$ [(Wm)$^{-1}$] & $\beta_2$ [ps$^2$/m] \\
\hline
530 & 0.186 & 323 & 58 & -2.12 \\
583 & 0.204 & 289 & 52 & -1.3 \\
620 & 0.217 & 275 & 49 & -0.89 \\
650 & 0.228 & 249 & 47 & -0.58 \\
690 & 0.242 & 234 & 44 & -0.27\\
820 & 0.287 & 190 & 39 &  0.45 \\
\end{tabular}
\caption{Nonlinear and GVD parameters used in the simulations of pulse propagation in silicon wire waveguides.}
\label{table : Si}
\end{table}

\vspace{0.5cm}
\noindent\textbf{Modeling of pulse propagation in InGaP photonic crystal and wire waveguides}. 
We use the following equation to simulate pulse propagation in InGaP waveguides~\cite{husko_free-carrier-induced_2016}:
\begin{equation}\label{eq : GNLSE3} \frac{\partial E(z,t)}{\partial z} = 
\left[-i\frac{\beta_2}{2}\frac{\partial^2}{\partial t^2} +\frac{\beta_3}{6}\frac{\partial^3}{\partial t^3}+i\frac{\beta_4}{24}\frac{\partial^4}{\partial t^4}-\frac{\alpha_\mathrm{l}}{2} - \left(\frac{\sigma}{2} + ik_0k_c\right)N_c\right]E
+ i\gamma |E|^2E - \alpha_{\mathrm{3PA}}|E|^4E.
\end{equation}
where $E(z,t)$ describes the slowly varying envelope of the field as a function of propagation distance $z$ and time $t$.
The equation differs from equation~\eqref{eq : GNLSE2}. The 2PA coefficient $\alpha_{\mathrm{2PA}}$ is absent as there is no two-photon absorption when pumping around the 1550~nm telecommunication wavelength. Instead the 3PA coefficient $\alpha_{3PA}$ is included. Also, the FCD parameter $k_c$ is considered independent of $N_c$ and we neglect the Raman effect as well as self-steepening. The carrier density is computed by solving
\begin{equation}\label{Eq : carriers3}
  \frac{\partial N_\mathrm{c}(z,t)}{\partial t} = \frac{2\alpha_{\mathrm{3PA}}}{3\hbar\omega_0 A_\mathrm{5eff}}|E(z,t)|^6-\frac{N_\mathrm{c}(z,t)}{\tau_\mathrm{c}}\,,
\end{equation}
where $A_\mathrm{5eff}$ is the effective area related to a fifth order process.
The values used in our simulations of photonic crystals are listed in table~\ref{table : InGaP}.
The normalized equation~\eqref{eq:normNlSE3} is linked to equation~\eqref{eq : GNLSE3} through the relations:
\begin{table}[h]
\centering
\begin{tabular}{l l l l l l l}
$N^2 = \frac{L_\mathrm{D}}{L_\mathrm{NL}}$ & $A = \frac{E}{\sqrt{P_0}}$ & $\zeta = \frac{z}{L_\mathrm{D}}$ &  $\tau = \frac{t}{T_0}$ & $\Gamma = \frac{\alpha_{\mathrm{3PA}}P_0}{\gamma}$ & $\kappa = \frac{2k_ck_0P_0T_0}{3\hbar\omega_0A_{\mathrm{5eff}}}$ \\
\end{tabular}
\end{table}

\noindent where  $L_\mathrm{D} = {T_0^2}/{|\beta_2|}$ and $L_\mathrm{NL}= {1}/{(\gamma P_0)}$ are the dispersion and nonlinear lengths as defined at the entrance of the waveguide.
\begin{table}[h]
\centering
\begin{tabular}{l l l l }
Parameter & PhCW & WW & Units\\
\hline
$\lambda_0$ & 1553 & 1550 & nm \\
$P_0$ & $5.9$ & 2.3 & W \\
$T_0$ & $1.25\times10^{-12}$ & $96.5\times10^{-15}$ & s \\
$\beta_2$ & $-4.7\times10^{-21}$ & $-6\times10^{-25}$ & s$^2$m$^{-1}$ \\
$\beta_3$ & $-5.5\times10^{-34}$ & NA & s$^3$m$^{-1}$ \\
$\beta_4$ & $6.7\times10^{-46}$ & NA & s$^4$m$^{-1}$ \\
$\alpha_\mathrm{l}$ & 920 & NA & m$^{-1}$ \\
$\sigma$ & $1.18\times10^{-20}$ & NA & m$^{3}$ \\
$k_c$ & $1.55\times10^{-26}$ & $5\times10^{-27}$ & m$^{3}$ \\
$\gamma$ & $2100$ & $450$ & (Wm)$^{-1}$ \\
$\alpha_{\mathrm{3PA}}$ & 7  & 0.28 & W$^{-2}$m$^{-1}$ \\
$A_{\mathrm{5eff}}$ & $2.3\times10^{-13}$ & $2.1\times10^{-13}$ & m$^{2}$
\end{tabular}
\caption{Physical parameters used in the simulations of pulse propagation in InGaP photonic crystal waveguides (PhCW) and wire waveguides (WW).
Note that the slow down factor $s$, inherent to the use of photonic crystal waveguides, is already taken into account in the listed values ($s=4.2$). NA: not applicable.}
\label{table : InGaP}
\end{table}
\pagebreak

\bibliographystyle{unsrt}
\bibliography{Ciret_SR_2018}

\section*{Acknowledgements}

This work was supported by the Belgian Science Policy Office (BELSPO); Interuniversity Attraction Poles Program Project Photonics (IAP 7-35); SP.G. and C.C. acknowledge the support from the Fonds de la Recherche Fondamentale Collective (FRFC) (PDR.T.1084.15). F.L acknowledges the support of the Fonds de la Recherche Scientifique (F.R.S.-FNRS, Belgium). C.H. was supported by the Alexei Abrikosov Fellowship. This work was performed, in part, at the Center for Nanoscale Materials, a U.S. Department of Energy Office of Science User Facility, and supported by the U.S. Department of Energy, Office of Science, under Contract No. DE-AC02-06CH11357. B.K. and F.L acknowledge funding from the European Research Council (ERC) under the European Union's Horizon 2020 research and innovation programme (grant agreement No 759483 \& 757800).

\section*{Author contributions statement}

\noindent C.C. carried out the experiments. SP.G. derived the normalized models. SP.G. and C.C. performed the simulations. B.K. and G.R. provided the waveguides. C.H. and SP.G. performed the analysis of the transition between channel and photonic crystal waveguides. All authors analyzed the results. F.L. supervised the project and wrote the manuscript with assistance from C.H. All authors reviewed the manuscript. 

\section*{Additional information}

\noindent\textbf{Supplementary figures} accompagnies this paper at...\\
\textbf{Competing financial interests:} The authors declare no competing interest\\
\textbf{Data availability statement:} The datasets generated during and/or analysed during the current study are available from the corresponding author on reasonable request.
\end{document}